\input phyzzx
\def\half{{1\over 2}}

\Pubnum={UUITP-08/97\cr 10A 97-06\cr}
\date={6 May 1997}
\titlepage
\title{Multidimensional Calogero systems from matrix models}
\bigskip
\author {Alexios P. Polychronakos\footnote\dagger
{poly@calypso.teorfys.uu.se}}
\address{Theoretical Physics Dept., University of Ioannina
\break 45110 Ioannina, Greece}
\andaddress{Theoretical Physics Dept., Uppsala University\break
S-751 08 Uppsala, Sweden}
\bigskip
\abstract{We show that a particular many-matrix model
gives rise, upon hamiltonian reduction, to a multidimensional
version of the Calogero-Sutherland model and its spin
generalizations. Some simple solutions of these models are
demonstrated by solving the corresponding matrix equations.
A connection of this model to the dimensional reduction of 
Yang-Mills theories to $0+1$-dimensions is pointed out. In 
particular, it is shown that the low-energy dynamics of D0-branes
in sectors with nontrivial fermion content is that of 
spin-Calogero particles.
}

\vfill
\endpage

\def\NP{{\it Nucl. Phys.\ }}
\def\PL{{\it Phys. Lett.\ }}

\def\PR{{\it Phys. Rev. \ }}
\def\PRL{{\it Phys. Rev. Lett.\ }}

\def\JMP{{\it J. Math. Phys.\ }}

\def\IJMP{{\it Int. Jour. Mod. Phys.\ }}

\def\LNC{{\it Lett. Nuovo Cimento \ }}

\REF\C{F.~Calogero, \JMP {\bf 10} (1969) 2191 and 2197; {\bf 12} (1971)
419; \LNC {\bf 13} (1975) 411; F.~Calogero and C.~Marchioro, \LNC
{\bf 13} (1975) 383.}
\REF\S{B.~Sutherland, \PR {\bf A4} (1971) 2019; {\bf 5} (1972) 1372;
\PRL {\bf 34} (1975) 1083.}
\REF\M{J.~Moser, {\it Adv. Math.} {\bf 16} (1975) 1.}
\REF\KKS{D.~Kazhdan, B.~Kostant and S.~Sternberg, {\it Comm. Pure Appl. Math.}
{\bf 31} (1978) 481.}
\REF\OP{M.A.~Olshanetskii and A.M.~Perelomov, {\it Phys. Rep.} {\bf 71} 
(1981) 314; {\bf 94} (1983) 6.}
\REF\GH{J.~Gibbons and T.~Hermsen, {\it Physica} {\bf D11} (1984) 337.}
\REF\W{S.~Wojciechowski, \PL {\bf A111} (1985) 101.}
\REF\HH{Z.N.C.~Ha and F.D.M.~Haldane, \PR {\bf B46} (1992) 9359.}
\REF\MP{J.~Minahan and A.P.~Polychronakos, \PL {\bf B302} (1993) 265.}
\REF\MPP{J.~Minahan and A.P.~Polychronakos, \PL {\bf B326} (1994) 288.}
\REF\CH{M.~Claudson and M.B.~Halpern, \NP {\bf B250} (1985) 689.}
\REF\BRR{M.~Baake, M.~Reinicke and V.~Rittenberg, \JMP {\bf 26} (1985) 1070.}
\REF\dHN{B.~de Witt, J.~Hoppe and H.~Nicolai, \NP {\bf B305} (1988) 545.}
\REF\DFS{U.H.~Danielsson, G.~Ferretti and B.~Sundborg, \IJMP {\bf A11}
(1996) 5463.}
\REF\KP{D.~Kabat and P.~Pouliot, \PRL {\bf 77} (1996) 1004.}
\REF\DKPS{M.R.~Douglas, D.~Kabat, P.~Pouliot and S.H.~Shenker, hep-th/9608024.}
\REF\BFSS{T.~Banks, W.~Fischler, S.H.~Shenker and L.~Susskind,
hep-th/9610043.}
\REF\GN{A.~Gorskii and N.~Nekrasov, {\it Theor. Math. Pnys.}
{\bf 100} (1994) 874.}

The quest for integrable (non-relativistic) particle systems in more 
than one spatial dimension is often frustrating. In general, no 
nontrivial such systems exist (with the exception of some isolated 
few-body cases), that is, systems with a non-quadratic
potential and which are not a repackaging of one-dimensional
degrees of freedom. In one
dimension, on the other hand, a whole class of integrable many-body
systems is known, namely the Calogero model and its various
generalizations (also known as Calogero-Sutherland-Moser systems) 
[\C-\M]. The question then is to what extent these models
remain solvable, if at all, in higher dimensions.

A particularly fruitful approach to Calogero-like systems is
though matrix models, in which the particle positions are regained
as the eigenvalues of some appropriate matrix [\KKS,\OP]. The 
integrability, as well as the solutions of the equations of motion, 
are simpler to obtain this way. It would seem, then, that this 
is the most promising route to systems of higher dimension. 
The purpose of this note is to show that, indeed, appropriate matrix
models give rise, under hamiltonian reduction, to multidimensional
many-body systems of the Calogero type. These matrix models
are not in general integrable, and therefore we do not expect the
corresponding particle systems to be integrable either. The 
hope is, nevertheless, that the (inherently simpler) matrix models
dynamics will allow for a better study of the dynamics of the 
particle systems.

The starting point will be a many-matrix model consisting of $d$
time-dependent hermitian $N \times N$ matrices $M_i$, $i=1, \dots d$,
which we will also represent as a vector of matrices ${\vec M}$.
The action will be the usual kinetic term for each matrix plus some 
potential invariant under simultaneous unitary conjugation of the 
matrices. The eigenvalues of $M_i$ then will be interpreted as the
$i$-component of the position vectors of $N$ particles moving
in flat $d$-dimensional space. For this interpretation to be
possible, however, the matrices must be simultaneously
diagonalizable, else there is no invariant association of the
$n$-th eigenvalue of the matrices as the coordinates of the
{\it same} particle. The eigenvalues of all $M_i$ can be 
simultaneously permuted
with a common unitary transformation, which corresponds to having 
identical particles. So we write the lagrangian
$$
L = \tr \left\{ \sum_i \half {\dot M}_i^2 + i \sum_{ij}
\half \Lambda_{ij} [ M_i , M_j ] \right\} - V({\vec M})
\eqn\L$$
where overdot stands for time derivative. $L_{ij} = -L_{ji}$
is an antisymmetric set of $d(d-1)/2$ hermitian matrices serving
as Lagrange multipliers for the commutativity constraint
between the $M_i$. The potential $V ({\vec M}) = V(U^{-1} {\vec M}
U)$ can be any real function of the $M_i$ invariant under simultaneous
unitary transformations of the $M_i$. The form $V = \tr V({\vec M})$, 
where $V({\vec x})$ is some real scalar function on $R^d$, will be 
assumed in what follows, which leads to an external potential 
$V({\vec x})$ for the particles. The harmonic 
oscillator potential $V({\vec x}) = \half \omega^2 {\vec x}^2$, 
corresponding to the matrix potential $V = \half \omega^2 \sum_i 
\tr M_i^2$, is the simplest example.

For a central potential the model is invariant 
under $SO(d)$ rotations $R_{ij}$:
$$
M_i \to R_{ij} M_j ~,~~
\Lambda_{ij} \to R_{ik} R_{jl} \Lambda_{kl}
\eqn\SOd$$
As a result, there is a conserved angular momentum
$$
J_{ij} = \tr \{ M_i {\dot M}_j - M_j {\dot M}_i \}
\eqn\J$$
Time translation invariance implies the conservation of energy
$$
E = \tr \left\{ \sum_i \half {\dot M}_i^2 + V(\{ M_i \} ) \right\}
\eqn\E$$
The invariance of \L\ under simultaneous conjugation of
all matrices by a time-independent unitary matrix implies the
existence of a conserved matrix ``angular momentum''
$$
K = i\sum_i [M_i , {\dot M}_i ]
\eqn\K$$
$K$ is a traceless hermitian matrix. Choosing it to have the 
``minimal'' form where all its eigenvalues are equal but one,
that is,
$$
K_{mn} = \ell ( \delta_{mn} - u_m^* u_n ) ~,~~ \sum_m |u_m |^2 = N
\eqn\minK$$
where $u$ is a fixed $N$-vector, will lead to Calogero dynamics for 
the eigenvalues of $M_i$, just as in the one-dimensional case,

The equations of motion are
$$
{\ddot M}_i + V_{,i} ({\vec M}) + i \sum_j [\Lambda_{ij} , M_j ] = 0
\eqn\EOM$$
plus the constraint
$$
[ M_i , M_j ] = 0
\eqn\con$$
The constraint implies that $M_i$ can all be diagonalized with
a common time-dependent unitary rotation $U(t)$:
$$
M_i = U^{-1} X_i U ~,~~ X_i = diag ( x_{i,1} , \dots x_{i,N} )
\eqn\diag$$
In terms of \diag\ the equations of motion acquire the form
$$
{\ddot X}_i + 2 [{\dot X}_i , A] + [X_i ,{\dot A}] +
\big[ [X_i ,A],A \big] + V_{,i} ({\vec X}) + i \sum_j [{\tilde \Lambda_{ij} }
, X_j ] = 0
\eqn\eqN$$
where ${\tilde \Lambda} = U \Lambda U^{-1}$ and $A = {\dot U} U^{-1}$
is the ``gauge potential'' generated by the time variance of $U$.

We now recall that the commutator of a diagonal matrix with
any matrix has zero diagonal elements, since
$$
[D,B]_{mn} = ( d_m - d_n ) B_{ij} ~,~~~{\rm if}~
D=diag ( d_1 , \dots d_N )
\eqn\diacom$$
Therefore, isolating the diagonal terms in \eqN, only the first,
fourth and fifth term contribute and we have
$$
{\ddot x}_{i,m} + \sum_n 2( x_{i,m} - x_{i,n} ) A_{mn} A_{nm} 
+ V_{,i} ({\vec x}_m ) = 0
\eqn\EMx$$
Plugging the form \minK\ and \diag\ in \K, on the other hand
we have
$$
i \sum_i ( x_{i,m} - x_{i,n} )^2 A_{mn} = \ell (\delta_{mn}
- {\tilde u}_m^* {\tilde u}_n )
\eqn\Kmn$$
where ${\tilde u} = U u$.
For $m \neq n$ and $m=n$ we obtain the relations for $A_{mn}$
and ${\tilde u}_m$, respectively
$$
A_{mn} = {i \ell {\tilde u}_m^* {\tilde u}_n \over 
( {\vec x}_m - {\vec x}_n )^2} ~,~~~ | {\tilde u}_m |^2 = 1
\eqn\Au$$
Plugging these in \EMx\ and calling $( x_{1,m} , \dots x_{d,m} ) =
{\vec x}_m$ we finally obtain
$$
{\ddot {\vec x}}_m -2 \ell^2 \sum_{n \neq m} { {\vec x}_m - 
{\vec x}_n \over ( {\vec x}_m - {\vec x}_n )^4} +
{\vec \nabla} V ( {\vec x}_m ) =0
\eqn\EMx$$
which is the equation of motion for the positions of particles
${\vec x}_m$ in an external potential $V({\vec x})$ and 
interacting through a $d$-dimensional two-body inverse square 
potential $\ell^2/x^2$, that is, a $d$-dimensional generalization
of the Calogero model. The key elements in the derivation are
that the Lagrange multiplier term does not influence the 
eigenvalue equations of motion and that the angular part 
reproduces the rotationally invariant $d$-dimensional 
inverse-square potential. Note that the energy \E\ and
angular momentum \J\ become the corresponding quantities
of the particle system in the constraint subspace, that is
$$
E = \sum_m \half {\dot {\vec x}}_m^2 + \sum_{m \neq n}
{\ell^2 \over ({\vec x}_m - {\vec x}_n )^2} + \sum_m V({\vec x}_m )
\eqn\Ep$$
$$
J_{ij} = x_i {\dot x}_j - x_j {\dot x}_i
\eqn\Jp$$

If the ``angular momentum'' $K$ is not in the ``minimal''
form \minK, it will enter the equations for the eigenvalues
in a nontrivial way and will give rise to multidimensional
generalizations of the `spin-Calogero' model [\GH-\MP]. To see 
this, we point out that the restriction of the Hamiltonian in
the constraint subspace $[M_i , M_j ]=0$ takes the form
$$
H = \sum_m \half {\dot {\vec x}}_m^2 + \sum_{m \neq n}
{{\tilde K}_{mn} {\tilde K}_{nm} \over ({\vec x}_m - 
{\vec x}_n )^2} + \sum_m V({\vec x}_m )
\eqn\Hcon$$
where ${\tilde K} = U K U^{-1}$. As usual, $\tilde K_{mn}$ 
Poisson-commute to the $SU(N)$ algebra and can be recast into 
internal degrees of freedom (``spin'') for the particles [\MPP]:
$$
{\tilde K}_{mn} = \sum_{a=1}^p S_m^a S_n^a
\eqn\spinK$$

To study the matrix equations of motion we specify to the
minimum nontrivial dimensions $d=2$ and to the rotationally
invariant harmonic external potential $V({\vec x})=\half
\omega^2 {\vec x}^2$. Defining the non-hermitian matrix
$M = M_1 + i M_2$, the equations of motion and constraint become
$$
{\ddot M} + [ \Lambda , M] + \omega^2 M =0 ~,~~
[ M , M^\dagger ] =0
\eqn\EMM$$
while the ``angular momentum'' K takes the form
$$
K = i [ M^\dagger , {\dot M} ]
\eqn\KM$$
in the constraint subspace. Solving the two-dimensional Calogero
model amounts to finding solutions of the above matrix equations
for $M$ with the form \minK\ for $K$.

The simplest possible class of solutions is the one with $\Lambda
=0$. It can be shown, however, that these solutions correspond
to linear motion of the particles and the system reduces to the
one-dimensional Calogero model. The solution to the equations of
motion is
$$
M = A e^{i\omega t} + B^\dagger e^{-i\omega t}
\eqn\solAB$$
where the matrices $A,B$, to satisfy the commutativity and ``angular
momentum'' constraints, must obey
$$
[A,B]=0 ~,~~ [A, A^\dagger ] = [B, B^\dagger ] = {K \over 2\omega}
\eqn\conAB$$
with $K$ as in \minK. In terms of the new matrices $Q=A+B^\dagger$
and $P=i\omega (A-B^\dagger )$ (representing the
position and velocity matrices at $t=0$) relations \conAB\ become
$$
[Q, Q^\dagger ] = [P, P^\dagger ] =0 ~,~~ [Q, P^\dagger ] = -iK
\eqn\conQP$$
This tells us that $P$ and $Q$, although non-hermitian, can each be 
diagonalized with a unitary rotation, with complex eigenvalues
(representing the initial positions and velocities of the particles
on the complex plane).
Choosing a basis where $Q$ is diagonal with eigenvalues $q_m$, 
we deduce from the last relation in \conQP\ that the matrix elements
of $P$ are
$$
P_{mn} =ip_m \delta_{mn} +{i\ell \over q_m^* - q_n^*} (1-\delta_{mn})
\eqn\Pmn$$
where we used $| u_m |=1$ in the $Q$-diagonal basis and further
chose the phases of the states such that $u_m =1$. From 
$[P, P^\dagger ] =0$ now we obtain
$$
{p_m - p_n \over q_m - q_n} = {\rm real} ~,~~~
\sum_{k \neq m,n} {1\over (q_m - q_k )(q_m^* - q_n^* )}
= {\rm real} 
\eqn\conP$$ 
By using the invariance of the equations by a shift of $Q$ and $P$
by a multiple of the unit matrix (which is related to the fact 
that the center-of-mass motion decouples from the relative motion)
we can always choose $q_1 = p_1 = 0$. Then the first relation above
implies that all $q_m$ are collinear (i.e., $q_m$/$q_n$=real) unless
$p_m = a q_m$ for some real $a$. The second relation, however, is 
satisfied only if the $q_m$ are collinear. By the first relation,
$p_m$ will also be collinear with $q_m$. Therefore, we see that the
one-dimensional Calogero model is included in the $\Lambda=0$ sector
of the general model. 

The other case in which the equations of motion
have an obvious solution is when $\Lambda$=constant. Choosing a basis
in which $\Lambda$ is diagonal, with (real) eigenvalues $\lambda_n$,
we have 
$$ 
{\ddot M_{mn}} + \omega_{mn}^2 M_{mn} = 0 ~,~~~ 
\omega_{mn} = \omega^2 + \lambda_m - \lambda_n 
\eqn\Mlambda$$ 
which has as solutions (we assume $\omega_{mn}^2 >0$)
$$ 
M_{mn} = A_{mn} e^{i\omega_{mn} t} + B_{mn}^\dagger 
e^{-i\omega_{mn} t}
\eqn\Mmn$$
The task of finding the most general $A_{mn}$, $B_{mn}$ which satisfy
the commutativity and ``angular momentum'' constraints is not trivial.
We demonstrate here a particularly simple solution, namely
$$
\eqalign{ &A_{mn} = A_m \delta_{mn} + a_n \delta_{m,n+1}~,~~ 
B_{mn} = B_n \delta_{mn}~, \cr
&|a_n|= a~,~~ \omega_{n+1,n+2} - \omega_{n,n+1}= {\ell \over a^2} \cr}
\eqn\ABmn$$
The last constraint for $\omega_{n,n+1}$ translates into $N-1$ 
algebraic equations for the $N-1$ variables $\lambda_n - 
\lambda_{n+1}$ (clearly $\Lambda$ can be shifted by any multiple of 
the unit matrix). It is obvious that the diagonal part $A_m$, $B_m$ 
represents a general motion of the decoupling center of mass. 
The eigenvalues of the off-diagonal part of $A$ are the $N$-th roots 
of $a_1 a_2 \cdots a_N$. So the off-diagonal part of $M$ has
eigenvalues
$$
z_m = x_m + i y_m = a e^{i({2\pi m \over N} + \omega_r t)} ~,~~~
\omega_r = \sum_n \omega_{n,n+1} =  \omega_{12} + (N-1) \ell /a^2
\eqn\solz$$
Therefore the relative motion is one in which the particles are
regularly positioned on a circle of radius $|a|$ and rotate with
constant angular velocity $\omega_r$.

The above model can be generalized to one with unitary matrices.
Omitting the details of the calculation, we simply state the
result. The lagrangian of the model is
$$
L = \tr \left\{ \sum_i \half R_i^2 {\dot U}_i^\dagger {\dot U}_i +
i \sum_{ij} \Bigl( \Lambda_{ij} [ U_i , U_j ] + 
\Lambda_{ij}^\dagger [ U_i^\dagger , U_j^\dagger ] \Bigr) \right\}
- V(\{ U_i \})
\eqn\LU$$
where again $V$ is some real conjugation-invariant potential. 
The Lagrange multiplier
matrices $\Lambda_{ij} = - \Lambda_{ji}$ are not hermitian, but 
the constraints arising from the variation of $\Lambda_{ij}$
and $\Lambda_{ij}^\dagger$ are compatible (in fact, equivalent)
for unitary $U_i$. The eigenvalues of $U_i$, written as
$\exp (i x_{i,m} /R_i )$ represent coordinates of particles on
a $d$-dimensional torus of radii $R_i$. Upon choosing the 
``angular momentum''
$$
i\sum_i R_i^2 [ U_i^\dagger , U_i ] = \ell (1 - u u^\dagger )
\eqn\KU$$
as before, the ${\vec x}_n$ move like particles on the torus
in an external potential $V({\vec x}_n )$ and interacting
through a periodic generalization of the $d$-dimensional
two-body inverse-square potential
$$
V( {\vec x}_n - {\vec x}_m ) = {\ell \over \sum_i \pi^2 R_i^2
\sin^2 {x_{i,m} - x_{i,n} \over \pi R_i}}
\eqn\Vsin$$
Similarly, the $d$-dimensional generalization of the 
inverse-sinh model can be obtained by taking $R_i \to i R_i$.

It is interesting to note a connection of the models presented
here with the matrix model obtained as the dimensional reduction of
(d+1)-dimensional Yang-Mills theory. In the $A_o =0$ gauge, this
model is essentially the kinetic part of the above
model plus a potential of the form
$$
\tr \sum_{ij} \half [ M_i , M_j ]^2
\eqn\MM$$
One can also consider the supersymmetric version of this model,
where there are also appropriate fermionic terms [\CH,\BRR]. In fact,
this model has also appeared as a regularization of the light-cone 
membrane action[\dHN]. For $d=9$ it describes the low-energy dynamics
of D-particles [\DFS-\DKPS] and has recently been proposed as a
matrix model description of M-theory in the large-$N$ limit [\BFSS].
This model can be thought of as our model where
a mass therm has been given to the Lagrange multiplier
matrices $\half \epsilon^2 \Lambda_{ij}^2$. Integrating out 
$\Lambda_{ij}$ (that is, solving their equation of motion) 
generates the 
above potential term \MM\ with a coefficient $1/\epsilon^2$.
For $\epsilon \to 0$, corresponding to our model, the strength of
the potential grows very large.
Therefore, our model can be though of as the low-energy limit
of the M-theory model, when the $M_i$ are at the minimum of
the potential \MM, that is, they commute. In the super-Yang-Mills
(M-theory) model there is also the Gauss law constraint
$$
\sum_i i [M_i , {\dot M}_i ] + \sum_i \Theta_i \Theta_i^T =0
\eqn\G$$
where $\Theta_i$ are Majorana-Weyl spinors, the superpartners of
$M_i$. The bosonic part of \G\ is just the ``angular momentum''
$K$. The fermionic part of \G\ provides real representations
of $SU(N)$, consisting of (the irreducible components of)
totally antisymmetric tensor products of adjoint representations
($\Theta$ transforms in the adjoint of $SU(N)$). Consequently,
\G\ constrains the ``angular momentum'' $K$ to be in one of these
representations, depending on the fermionic sector of the model.
Therefore we conclude that the low-energy motion of the eigenvalues
of $M_i$ (that is, the coordinates of the so-called D0-branes
in M-theory) is the one of particles in the $d$-dimensional 
spin-Calogero model ($d=9$ for M-theory),
where the spin degrees of freedom of the particles, which
participate in the dynamics, are determined by the fermionic
state of the model through \G\ and \spinK. The physical 
implications of this fact are left for future study.

We conclude with some remarks. Clearly, there are a lot of
unanswered questions here. Firstly, the physical meaning of the
Lagrange multiplier matrices $\Lambda_{ij}$ and their role
at classifying the types of solutions are yet to be understood.
The matrix equations of motion have barely been touched in the 
general case, and their solutions are unknown. Even in the
case $\Lambda$=constant the general solution has not been fully
studied. Generalizations of these models involving the Weierstrass
function potential could be sought, where instead of a matrix model
one would have to deal with an appropriate topological model [\GN].
In fact, it would be interesting to consider what type of model
would give rise to Calogero-type dynamics on a manifold of more
general geometry and/or topology. Finally, the quantization of the
model, being a constrained system, is a subject of investigation.

\refout
\end